\documentclass{article}

\usepackage{arxiv}

\usepackage[utf8]{inputenc} % allow utf-8 input
\usepackage[T1]{fontenc}    % use 8-bit T1 fonts
\usepackage{hyperref}       % hyperlinks
\usepackage{url}            % simple URL typesetting
\usepackage{booktabs}       % professional-quality tables
\usepackage{amsfonts}       % blackboard math symbols
\usepackage{nicefrac}       % compact symbols for 1/2, etc.
\usepackage{microtype}      % microtypography
\usepackage{lipsum}
\usepackage{graphicx}
\usepackage{natbib}
\usepackage{algorithm}% http://ctan.org/pkg/algorithms
\usepackage{algpseudocode}% http://ctan.org/pkg/algorithmicx
\usepackage{amsmath}
\usepackage{multirow} 
\graphicspath{ {./images/} }

\begin{document}

\title{Rashomon effect in Educational Research: Why More is Better Than One for Measuring the Importance of the Variables?}
\date{} %do not delete this, it suppresses insertion of the date

\author{
 Jakub Kuzilek \\
  Humboldt-Universität zu Berlin\\
  Berlin, Germany \\
  \texttt{jakub.kuzilek@hu-berlin.de} \\
  %% examples of more authors
   \And
   Mustafa Çavuş \\
    Eskisehir Technical University\\
    Eskişehir, Turkey \\
  \texttt{mustafacavus@eskisehir.edu.tr}
  }

\maketitle
\begin{abstract}
This study explores how the Rashomon effect influences variable importance in the context of student demographics used for academic outcomes prediction. Our research follows the way machine learning algorithms are employed in Educational Data Mining, focusing on highlighting the so-called Rashomon effect. The study uses the Rashomon set of simple-yet-accurate models trained using decision trees, random forests, light GBM, and XGBoost algorithms with the Open University Learning Analytics Dataset. We found that the Rashomon set improves the predictive accuracy by 2-6\%. Variable importance analysis revealed more consistent and reliable results for binary classification than multiclass classification, highlighting the complexity of predicting multiple outcomes. Key demographic variables \texttt{imd\_band} and \texttt{highest\_education} were identified as vital, but their importance varied across courses, especially in course DDD. These findings underscore the importance of model choice and the need for caution in generalizing results, as different models can lead to different variable importance rankings. The codes for reproducing the experiments are available in the repository: \url{https://anonymous.4open.science/r/JEDM_paper-DE9D}.

%This document serves both as JEDM submission instruction and as a template file.  This is the abstract. It should contain from 100 to 300 words. Authors are encouraged to share the code, data, or intermediate results behind their submissions. While links to code can be included in footnotes in the article, authors are encouraged to include a sentence at the end of the abstract to draw the attention of readers, something like: The code is available at https://kwbln.github.io/code3 and the  data and at: https://xyz.github.io/jedm23. \\ %Keep \\ for spacing to keywords

{\parindent0pt
\textbf{Keywords:} Rashomon effect, student success prediction, student demographics, variable importance, responsible machine learning
}
\end{abstract}

\section{Introduction}

The fields of educational data mining (EDM) and learning analytics (LA) focus on the exploration and use of student and study-related data for modelling the learning processes, success, and learning to support teaching and learning on different educational levels \citep{siemens2012learning}. Both fields share many aspects, and one is the dependency on machine learning (ML) models for modelling the underlying phenomena \citep{siemens2012learning}. In the past decade, ML, enabled by the growth of the data collected within education, provided vast amounts of insights into the studied problems \citep{papamitsiou2014learning} ranging from the predictive modelling of student outcomes using various student and study-related characteristics (e.g. \cite{arnold2012course,Kennedy2015, davis2016gauging}) to learner modelling \citep{Abyaa2019} or knowledge tracing \citep{abdelrahman}. 

Most studies train models and use them in line with good machine learning standards following the so-called data science workflow \citep{wickham2023}, which contains a series of steps to produce reproducible and reliable results. The workflow starts with importing and transforming the data into a form suitable for use by ML algorithms. Next, the ML algorithm(s) is applied to produce a model approximating the studied phenomenon \citep{murphy_2022}. The process in which the models are tuned and the best-fitting model is selected often involves using cross-validation techniques and several measures of the quality of the fit \citep{Raschka2018ModelEM}. The model training involves several questions, which influence the resulting model \citep{Rudin2024}. For example: What is the precision-recall tradeoff? Which algorithm is the most suitable for the given data? Is the final model interpretable? Can we estimate the variable importance? As stated, these questions heavily influence the experiment results and the selection of the best-fitting model. At the end of this process, researchers get the model of the studied phenomena, which is the best approximation given the dataset and the used training algorithm.

It has been noted over two decades ago that real-world data allows many approximately equally good models \citep{breiman2001statistical}. The phenomenon has been called the Rashomon effect after the Kurosawa movie Rashomon \citep{kurosawa1950}, in which four different witnesses spotted a murder and described the event differently, showing that there is no single truth. The Rashomon effect thus unlocks the possibilities for exploring the information from the sets (families) of the predictive models while impacting issues like fairness and explainability, model simplicity, or reliable variable importance \citep{Rudin2024,biecek_et_al_2024}. The effect often manifests strongly with the data generated by the noisy processes  \citep{semenova2023}, which is the case of educational data, since the recorded data approximate complex learning processes. The extensive Rashomon effect also implies the existence of simple-yet-accurate models \citep{Rudin2024}. The Rashomon effect has many implications for researchers tackling ML problems. In our particular case - variable importance since one cannot assume that variables important for one model would be significant to every equally-performing model. 

As shown by previous research, learner characteristics such as age, gender, etc impact the learning outcomes of the individual learner \citep{Wladis2014, Romero2013}. Yet, the importance of such indicators is also influenced by the learners' attitude toward the learning itself \citep{Kuzilek2015}. Generally, researchers use as many variables as possible to improve learning outcome prediction (e.g. \cite{Arnold2012, Kennedy2015, Sharabiani2014, Shehata2015, Kent2017}). Only a few studies have examined the impact of demographic characteristics on learners' performance using variable importance. For example, \cite{rizvi_et_al_2019} investigated the importance of demographic variables in the decision tree models trained using the Classification and Regression Tree algorithm \citep{breiman_et_al_1984} on a sample consists four courses from the Open University Learning Analytics Dataset (OULAD) \citep{Kuzilek2017}. The study is representative of the current machine learning paradigm, which tends to derive results from using only the best model. Due to the nature of the learning, there can be multiple well-performing predictive models for a given dataset, and the important variables in one model may not necessarily be important in others \citep{donnelly_et_al_2023}. \citet{rodolfa_et_al_2020} highlighted the importance of the Rashomon effect and demonstrated that multiple well-performing models can lead to different demographic disparities.

Our study focuses on showing how the Rashomon effect affects the variable importance calculation following the approach introduced by \cite{rizvi_et_al_2019}. Our approach replaces a single model - a decision tree, with a set of simple-yet-accurate tree models, called the Rashomon set, to show how the used model set influences the variable importance, ultimately reducing the dependency on single model selection. The task is to predict the final course result given the demographic variables. We formulated the following research questions:
\begin{itemize}
    \item[RQ1] \textit{How does the choice of training algorithm influence the predictive performance of the final model?};
    \item[RQ2] \textit{Are there any differences between the multiclass and binary classification tasks in terms of variable importance?};
    \item[RQ3] \textit{How does the choice of models affect the variable importance?}
    
\end{itemize}

The paper is organised as follows: First, the state of the art is analysed; next, the data and methods are described, followed by results and their discussion. Finally, the paper concludes the proposed study and summarises the findings and some limitations.

\section{Related work}

The analysis of student data experienced a massive boom in the past decade due to the increased capabilities in measuring, storing, and processing vast amounts of data. One of the prominent tasks that received the focus of many researchers is academic performance prediction \citep{Papamitsiou2014}, because it is a vital part of learning crucial to the learners, teachers and universities. The performance itself is reflected by various measures, starting from the individual learner exam results and ending with the overall success of students in different study programmes. Due to this fact, the problem of predictive modelling has been extensively studied (e.g. \cite{Arnold2012,Kennedy2015,Sharabiani2014,Shehata2015,Kent2017,Wladis2014} and many others), and there exist systems which employ the predictive models in day-to-day teaching \citep{Arnold2012,Kuzilek2015,klerkx2017learning}. 

The analysis of learner-related data has been shifting from demographics to much wider datasets using data collected from Virtual Learning Environments (VLEs), which has been proven to be the source of helpful information \citep{Arnold2012,Kennedy2015}. Yet, the student demographics provide useful information about the given population \citep{richardson2009academic}. The research on the demographics influence academic success in traditional or online learning, student behaviour, and performance estimation is vast \citep{papamitsiou2014learning, balaji2021contributions}. The most studies use several well-know characteristics including gender \citep{kizilcec2017self}, age \citep{boyte2018age}, education level \citep{Wladis2014, kizilcec2017self}, geographical location \citep{bayeck2018influence}, and socio-economic status \citep{Wladis2014}. Student demographics also play a critical role in student success estimation with no VLE-related data available \citep{Sharabiani2014,Shehata2015}. 

To further explore the importance of demographics, the variable importance methods can be used \citep{molnar_et_al_2020}. Those methods explore the relevance of each characteristic to the modelled phenomena. Only a few examples employing the variable importance techniques in the educational domain exist. For example, \cite{rizvi_et_al_2019} investigated the importance of demographic variables using the Classification and Regression Tree algorithm. The characteristics included all demographical characteristics from the OULAD dataset (age, gender, geographical region, socioeconomic status, disability and highest education). The study uncovered a strong relationship between the region, socioeconomic status and academic outcomes. More recently \cite{harif2024predictive} employed the Support Vector Machines combined with Random Forest and Recursive Feature Elimination with Cross-Validation to predict student outcomes in university courses. The feature elimination step identified the importance of the geographical location together with the previous education outcomes. 

Finally, it is essential to point out that most research employs only a standard machine learning cycle, resulting in one model used for the target phenomena. Yet, the important variables in one model may not necessarily be important in others \citep{donnelly_et_al_2023}. \citet{rodolfa_et_al_2020} highlighted the importance of the Rashomon effect and demonstrated that multiple well-performing models can lead to different demographic disparities. \citet{fisher_et_al_2019} introduced the concept of the model class reliance as the range of variable importance values across the multiple well-performing models. Similarly, \citet{dong_and_rudin_2020} introduce Variable Importance Clouds, designed to offer insights into the hidden dynamics of these models. These concepts provide a broader perspective on variable importance by accounting for the variability in importance across multiple good models rather than relying on a single model’s interpretation \citep{rudin_et_al_2022}.\\
%%%%%%%%%%%%%%%%%%%%%%%%%%%%%%%%%%%%%%%%%%%%%%%%%%%%%%%%%%%%%%%%%%%%%%%%%%%%%
\section{Methods}

In this section, we first define the Rashomon effect and the related concepts. The permutational variable importance is given, and lastly, we summarize the dataset used in modelling and the experimental design. 
%%%%%%%%%%%%%%%%%%%%%%%%%%%%%%%%%%%%%%%%%%%%%%%%%%%
\subsection{Rashomon Effect}
Let $ D = \{(x_i, y_i)\}_{i=1}^n $ is a dataset of $n$ observations, where $ x_i \in X $ represents observations from the feature set and $ y_i \in Y $ represents response vectors. The model space is denoted by $ F = \{f \mid f: X \to Y\} $. The objective is to find a model $ f \in F $ that minimizes the expected loss. The expected loss is represented as $ E[L(f)] $, where $ L: F \to \mathbb{R} $ is the loss function. The expected loss is approximated by the empirical loss based on the dataset $ D $. The model with the minimum empirical loss, referred to as the reference model $ f_R $, is given by:
\begin{equation}
    f_R = \arg \min_{f \in F} E[L(f)].    
\end{equation}

\noindent \textbf{Rashomon Set}: For a given loss function $ L $, reference model $ f_R $, and a Rashomon parameter $ \epsilon > 0 $, the Rashomon set $ \mathcal{R}_{L,\epsilon}(f_R) $ is defined as:
\begin{equation}
    \mathcal{R}_{L,\epsilon}(f_R) = \{f \in F \mid E[L(f)] \leq E[L(f_R)] + \epsilon\}.
\end{equation}

\noindent Since accessing the models space (aka all possible models) in $ F $ is infeasible, we focus on an empirical model space $ \hat{F} \subset F $ and the corresponding empirical Rashomon set:
\begin{equation}
    \hat{\mathcal{R}}_{L,\epsilon}(f_R) = \{f \in \hat{F} \mid E[L(f)] \leq E[L(\hat{f}_R)] + \epsilon\}.    
\end{equation}

\noindent The model space refers to all possible models that minimize the expected loss, and the Rashomon set is a subset of the model space that consists of highly and similarly performant models belonging to the model space. On the other hand, the \textbf{Rashomon Set Size} quantifies the number of models in this set, indicating the extent of the Rashomon effect. A larger size suggests many viable models, implying a stronger Rashomon effect. 
%%%%%%%%%%%%%%%%%%%%%%%%%%%%%%%%%%%%%%%%%%%%%%%%%%%
\subsection{Permutational Variable Importance}

Permutational Variable Importance (PVI) is a model-agnostic technique used to evaluate the importance of individual variables in a predictive model \citep{breiman_2001}. The basic idea is to measure how much the model's predictive performance decreases when the values of a single variable are randomly shuffled, breaking the relationship between that variable and the target variable. This method can be applied to any model, including complex non-linear models like Random Forests or Neural Networks, making it a versatile tool in model interpretation.

\begin{algorithm}
\caption{Procedure for Calculating Permutation Variable Importance}
\begin{algorithmic}[1]
    \State \textit{Train the Model}: Train the predictive model on the original dataset.
    
    \State \textit{Calculate Baseline Performance}: Evaluate the model's performance on a validation set to obtain a baseline score $ S_{\text{baseline}} $. This could be accuracy, AUC, mean squared error, or any other relevant metric depending on the task.
    
    \For{each variable $ j $ in the dataset}
        \State \textit{Permute the Variable Values}: Randomly shuffle the values of variable $ j $ across all instances in the validation set.
        
        \State \textit{Re-evaluate the Model}: Re-evaluate the model's performance on the validation set, yielding $ S_{\text{permute}_j} $.
        
        \State \textit{Compute Variable Importance}: Calculate the importance score for variable $ j $ as the average decrease in performance over $ M $ repetitions:
        \[
        I_j = \frac{1}{M} \sum_{i=1}^{M} (S_{\text{baseline}} - S_{\text{permute}_j})
        \]
    \EndFor
    
    \State \textbf{Return:} Return the importance scores $ I_j $ for all variables.
\end{algorithmic}
\end{algorithm}

The larger the drop in performance, the more important the variable is to the model's predictions. Variables that do not contribute significantly to the model will have little or no effect on performance when permuted, resulting in a low importance score. This method provides an intuitive and interpretable way to rank variables by their contribution to the model's predictive accuracy, and it is beneficial for understanding black-box models.

\subsection{Variable Importance Order Discrepancy}

The Variable Importance Order Discrepancy (VIOD) measures the dissimilarity between the variable importance orders of models in the Rashomon set using Kendall’s $ \tau $ correlation coefficient \citep{kendall_1938}. The VIOD is defined as:
\begin{equation}
    V_\epsilon(f_R) = \max_{f \in \mathcal{R}_{L,\epsilon}(f_R)} \tau(f_R, f)
\end{equation}

\noindent where $ V_\epsilon(f_R) $ represents the maximum discrepancy in variable importance orders between the reference model $ f_R $ and other models in the Rashomon set \citep{cavus_and_biecek_2024}. The VIOD captures the extent of variation in variable prioritization across different models compared to the reference model. It takes the values between $-1$ and $1$. Its lower values signify higher consistency in variable importance rankings, while the higher values indicate greater variability.

%%%%%%%%%%%%%%%%%%%%%%%%%%%%%%%%%%%%%%%%%%%%%%%%%%%
\subsection{Dataset}

The Open University Learning Analytics Dataset (OULAD) is publicly accessible \citep{kuzilek_et_al_2017} from the Open University, which is the largest distance education provider in the UK. Open University courses typically span nine months and involve multiple assignments and a final exam. A student's success in a course is determined by performance on assessments and the final exam. The OULAD dataset contains information on student demographics such as \texttt{age\_band}, \texttt{disability}, \texttt{education}, \texttt{gender}, \texttt{imd\_band}, and \texttt{region}, assessment outcomes, and click-stream data from interactions with a learning management system similar to Moodle. We focused on courses AAA, BBB, DDD, and EEE, completed by 309, 2646, 2017, and 766 students, respectively. For the sake of simplicity, we will be referring to them using single characters only (e.g. A for AAA). The final course outcome, categorized as \texttt{Distinction}, \texttt{Pass}, or \texttt{Fail}, was used as the target variable for model training. We use it as such for a multiclass classification setting. In a binary classification task, students with \texttt{Distinction} were grouped with those who earned a \texttt{Pass}, creating a binary classification between \texttt{Pass} and \texttt{Fail}. Our analysis concentrated on demographic data. Thus, the final dataset included six variables summarizing students' demographics. Table~\ref{tab:dataset} presents detailed descriptions of the variables used in modelling.

\begin{table}[h]
    \centering
    \caption{Description of selected dataset variables}
    \label{tab:dataset}
    \begin{tabular}{lp{6cm}p{6cm}}\toprule
    Variable        & Description               &  Values \\\midrule
    \texttt{age\_band}       & a band of student’s age  & $\{0-35, 35-55, 55+\}$ \\
    \texttt{disability}      & whether the student has declared a disability   & $\{\text{TRUE}, \text{FALSE}\}$  \\
    \texttt{education}       & the highest student education level on entry to the module presentation  & $\{$Lower Than A Level, A Level or Equivalent, HE Qualification, Post Graduate Qualification$\}$ \\
    \texttt{gender}          & student’s gender  & F, M \\
    \texttt{imd\_band}       & the IMD band of the place where the student lived during the
module presentation  & $\{0-10\%, 10-20\%, 20-30\%, 30-40\%, 40-50\%, 50-60\%, 60-70\%, 70-80\%, 80-90\%, 90-100\%\}$  \\
    \texttt{region}          & the geographic region, where the student lived while taking the
module presentation  & $\{$East Anglian Region, East Midlands Region, Ireland, London Region, North Region, North Western Region, Scotland, South East Region, South Region, South West Region, Wales, West Midlands Region, Yorkshire Region$\}$ \\
    \texttt{final\_result}   &   & $\{$Pass, Fail$\}$  \\\bottomrule
    \end{tabular}
\end{table}
%%%%%%%%%%%%%%%%%%%%%%%%%%%%%%%%%%%%%%%%%%%%%%%%%%%%%%%%%%%%%%%%%%%%%%%%%%%%%
\subsection{Experimental Design}

\begin{figure}[h]
    \centering
    \includegraphics[scale = 0.52]{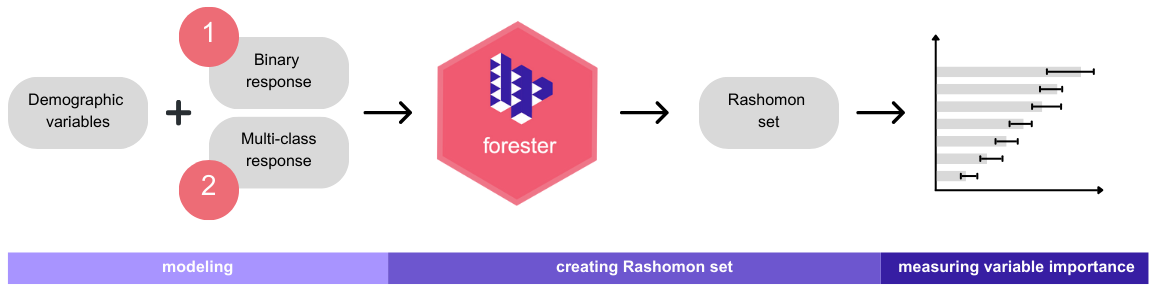}
    \caption{The workflow of the experimental design}
    \label{fig:design}
\end{figure}

In this section, the experimental design is described, which is conducted to answer the research questions considered in this paper. The approach, illustrated in Figure~\ref{fig:design}, involves the following steps: (1) building the model structure with binary and multiclass responses to investigate the difference between the type of response variable, (2) creating the Rashomon set using \texttt{forester} which is a tree-based AutoML tool \citep{ruczynski_and_kozak_2024}, and (3) measuring and summarizing the importance of the demographical variables of the models in the Rashomon set. 

\noindent \textbf{Modelling.} We are interested in the difference between the type of response variable used in student success prediction models regarding variable importance. Thus, binary and multiclass type of \texttt{final\_result} is used as the response variable with the levels \texttt{Fail} - \texttt{Pass} (merging the levels \texttt{Pass} and \texttt{Distinction}), and \texttt{Fail} -  \texttt{Pass} - \texttt{Distinction}, respectively. The demographic variables are used as predictor variables in both of the models (see Table~\ref{tab:dataset}).\\

\noindent \textbf{Creation Rashomon Set.} We used \texttt{forester} consisting of models such as decision tree \citep{breiman_et_al_1984}, random forest \citep{breiman_2001}, lightGBM \citep{ke_et_al_2017}, and XGBoost \citep{chen_and_carlos_2016} tuned by Random Search \citep{bergstra_et_al_2012} and Bayesian optimization \citep{snoek_et_al_2012} to create a Rashomon set because it allows controlling the set size through its Bayesian optimization parameters. We created a model space of $424$ tree-based models by setting these parameters as \texttt{bayes\_iter} $= 30$ and \texttt{random\_evals} $= 200$. Then, the Rashomon sets of the courses were created for the Rashomon parameter $\epsilon = 0.05$. \\

\noindent \textbf{Measuring Variable Importance.} The Permutational Variable Importance (PVI) is used to measure the importance of each variable in this phase. These values are calculated for each of the models in the Rashomon set. The distributions of PVI of the models are given in Figure~\ref{fig:pvi}. Moreover, the Variable Importance Order for each model in the Rashomon set is visualized in terms of their distributions given in Figure~\ref{fig:kendall}.

%%%%%%%%%%%%%%%%%%%%%%%%%%%%%%%%%%%%%%%%%%%%%%%%%%%
\section{Results}

In this section, the results of the experiments are evaluated in three directions: (1) the comparison of model performance in the model space and Rashomon set, (2) the distribution of the permutational variable importance, (3) the distribution of the VIOD values for the tasks and courses. \\

\noindent \textbf{The comparison of model performance in the model space and Rashomon set.} The mean and standard deviation of the model's performances in the model space and the Rashomon set, along with the Rashomon set size for each course and classification setup, is given in Table~\ref{tab:perf}. 

\begin{table}[]
    \centering
    \caption{The summary of the model accuracy in the model space and the Rashomon set (mean $\pm$ sd)}
    \label{tab:perf}
    \begin{tabular}{lcccc}\toprule
       Setup & Course   & Model space          & Rashomon set       & Rashomon set size \\\midrule
       \multirow{4}{*}{binary}& A        & 0.8085 $\pm$ 0.193   & 0.8650 $\pm$ 0.009 & 381 \\
       & B        & 0.6755 $\pm$ 0.102   & 0.7051 $\pm$ 0.005 & 384\\
       & D        & 0.6246 $\pm$ 0.065   & 0.6562 $\pm$ 0.008 & 232\\
       & E        & 0.6883 $\pm$ 0.136   & 0.7373 $\pm$ 0.007 & 354\\\midrule
       \multirow{4}{*}{multiclass}& A      & 0.7513 $\pm$ 0.173 & 0.8039 $\pm$ 0.008 & 374\\
       & B      & 0.5342 $\pm$ 0.101 & 0.5676 $\pm$ 0.010 & 372\\
       & D      & 0.5457 $\pm$ 0.101 & 0.5851 $\pm$ 0.011 & 261\\
       & E      & 0.5279 $\pm$ 0.116 & 0.5792 $\pm$ 0.005 & 316\\\bottomrule
    \end{tabular}
\end{table}\vspace{3mm}

For each setup, the average accuracy and standard deviation (mean $\pm$ sd) of the model space are provided, along with the corresponding values for the Rashomon set. Course A exhibits the highest accuracy in the binary classification setup in the model space ($0.8085 \pm 0.193$) and the Rashomon set ($0.8650 \pm 0.009$), with $381$ models in the Rashomon set. Course B follows with a model space accuracy of $0.6755 \pm 0.102$ and Rashomon set accuracy of $0.7051 \pm 0.005$, containing $384$ models. For Courses D and E, the Rashomon set also shows slight improvements over the model space, with $232$ and $354$ models, respectively, indicating moderate consistency in the results across these courses. Course A again demonstrates strong performance in the multiclass setup, achieving $0.7513 \pm 0.173$ accuracy in the model space and $0.8039 \pm 0.008$ in the Rashomon set, comprising $374$ models. Course B’s accuracy in the model space is $0.5342 \pm 0.101$, improving to $0.5676 \pm 0.010$ in the Rashomon set, with a total of $372$ models. Courses D and E present lower accuracies, with relatively smaller Rashomon sets, indicating that fewer models meet the performance criteria in these cases. The Rashomon set analysis suggests multiple models with comparable performance, especially for Courses A and B, across both setups. This redundancy implies flexibility in choosing models without compromising accuracy, though the number of such viable models varies by course and classification type.\\

\begin{figure}[h]
    \centering
    \includegraphics[scale = 0.345]{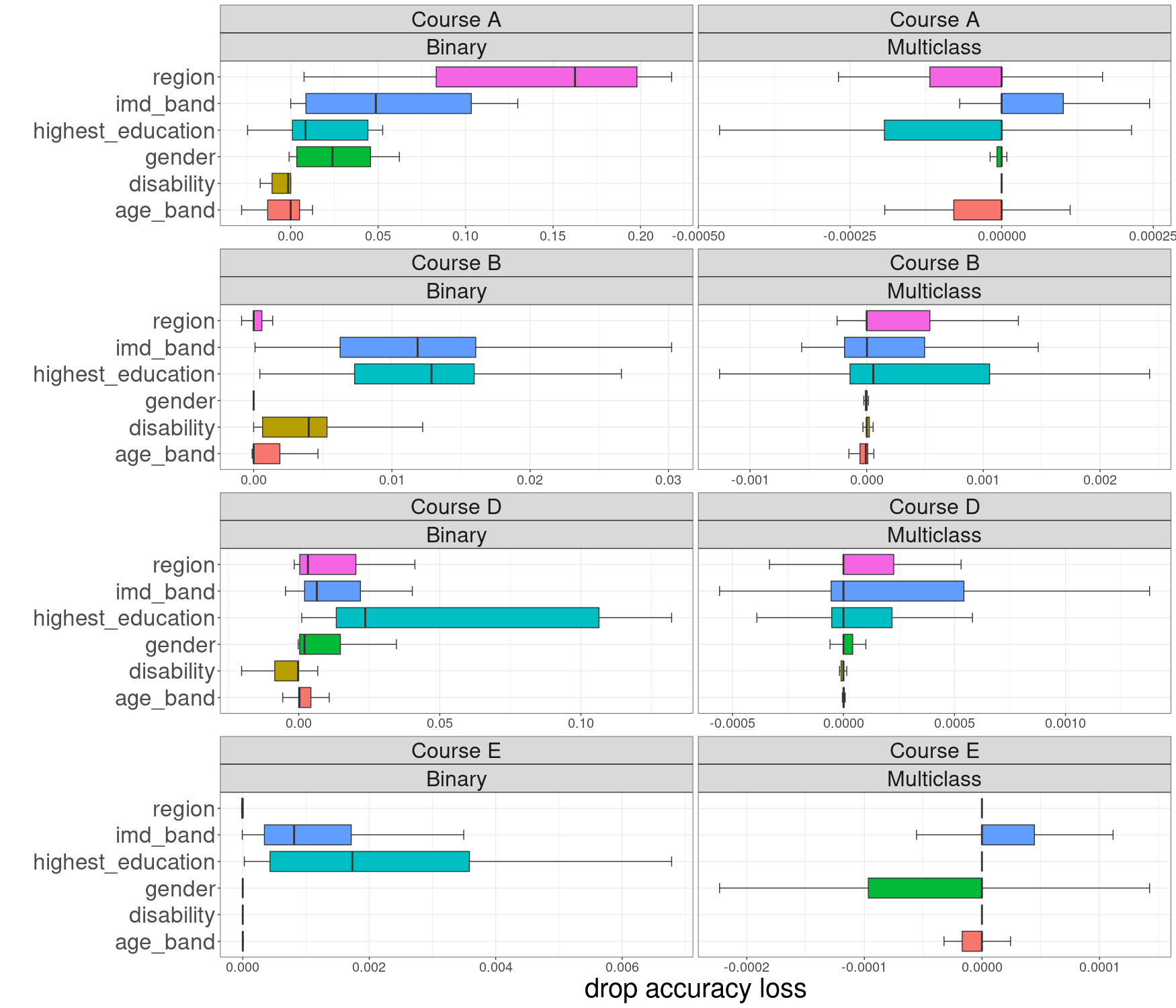}
    \caption{Permutational variable importance distributions of the models for classifying the \texttt{final\_result}}
    \label{fig:pvi}
\end{figure}

\noindent \textbf{The distribution of the permutational variable importance.} Figure~\ref{fig:pvi} presents the drop accuracy loss values for various demographic variables across five courses A, B, D, and E under binary and multiclass response settings. The drop accuracy loss indicates the reduction in predictive performance when a particular variable is excluded from the model, suggesting the importance of that variable for the model's accuracy.

In Course A, the binary model identifies gender as the most important variable, followed by \texttt{imd\_band} and \texttt{region}, with notable drops in accuracy when these variables are removed. In contrast, the multiclass model for Course A highlights \texttt{region} and \texttt{imd\_band} as the most important variables, while other variables such as \texttt{disability} and \texttt{age\_band} make minimal contributions. For Course B, the binary and multiclass models show that \texttt{imd\_band} and \texttt{highest\_education} play a crucial role in maintaining accuracy. Although \texttt{gender} is somewhat relevant in the binary setting, its impact is less evident in the multiclass model. In Course D, the \texttt{highest\_education} variable emerges as the dominant factor, with significant accuracy loss observed in both binary and multiclass models when this variable is excluded. Other variables, such as \texttt{gender} and \texttt{imd\_band}, contribute somewhat. Finally, the binary model for Course E emphasizes \texttt{highest\_education} and \texttt{imd\_band} as important variables. In contrast, the multiclass model continues to show the importance of \texttt{imd\_band} with only minor contributions from other variables. Notably, the multiclass models generally exhibit smaller drop accuracy loss values than the binary models, indicating that no single variable overwhelmingly influences performance. The results suggest that \texttt{imd\_band} and \texttt{highest\_education} are consistently important across multiple courses, particularly in binary models. The multiclass models, on the other hand, exhibit smaller drop accuracy loss values, indicating that no single variable has an overwhelming influence on performance. A notable exception is Course D, where the number of important models and the accuracy impact are limited, which may explain the lower performance observed for this course. This suggests that the diversity and size of the Rashomon set and the presence of important variables play a significant role in determining model accuracy across courses and response types.\\

\noindent \textbf{The VIOD for the tasks and courses.} Table~\ref{tab:viod} presents the variability in the importance of variables across binary and multiclass models for predicting student success in the courses. The Variable Importance Order Discrepancy (VIOD) metric captures the maximum dissimilarity in the order of variable importance between the highest-performing model and other models within the same setup (binary or multiclass). A higher VIOD value close to the $1$ indicates greater consistency in variable importance rankings, whereas a lower value close to the $-1$ reflects more variability. 

The binary models generally exhibit low VIOD values, with scores ranging from $-0.066$ to $0.200$, indicating stable variable importance across models. Courses A and E display nearly identical VIOD values of $-0.066$, suggesting a high degree of agreement among the models regarding which variables are important. Course B shows slightly more variability with a VIOD of $0.066$, but the dissimilarity remains minimal. In contrast, Course D demonstrates the highest dissimilarity among the binary models, with a VIOD of $0.200$, reflecting greater inconsistency in variable rankings across the different models used.

In the multiclass setup, the models show substantially higher VIOD values, ranging from $-0.600$ to $-0.866$, pointing to less stable variable importance. Courses A and B share a VIOD value of $-0.600$, indicating moderate variability in the importance order among the top-performing models. Course E shows even higher variability, with a VIOD value of $-0.733$, suggesting that the models identify different variables as important. Course D stands out with the highest VIOD score of $-0.866$, indicating that the models exhibit significant differences in their rankings of variable importance, which could reflect the difficulty of predicting outcomes consistently for this course.\\

\begin{table}[]
    \centering
    \caption{Variable importance order discrepancy for courses}
    \label{tab:viod}
    \begin{tabular}{ccr}\toprule
       Setup & Course & VIOD    \\\midrule
       \multirow{4}{*}{binary}     & A      & -0.066  \\
             & B      & 0.066     \\
             & D      & 0.200     \\
             & E      & -0.066     \\\midrule
       \multirow{4}{*}{multiclass}      & A      & -0.600     \\
             & B      & -0.600     \\
             & D      & -0.866     \\
             & E      & -0.733     \\\bottomrule
    \end{tabular}
\end{table}

\begin{figure}[h]
    \centering
    \includegraphics[scale = 0.345]{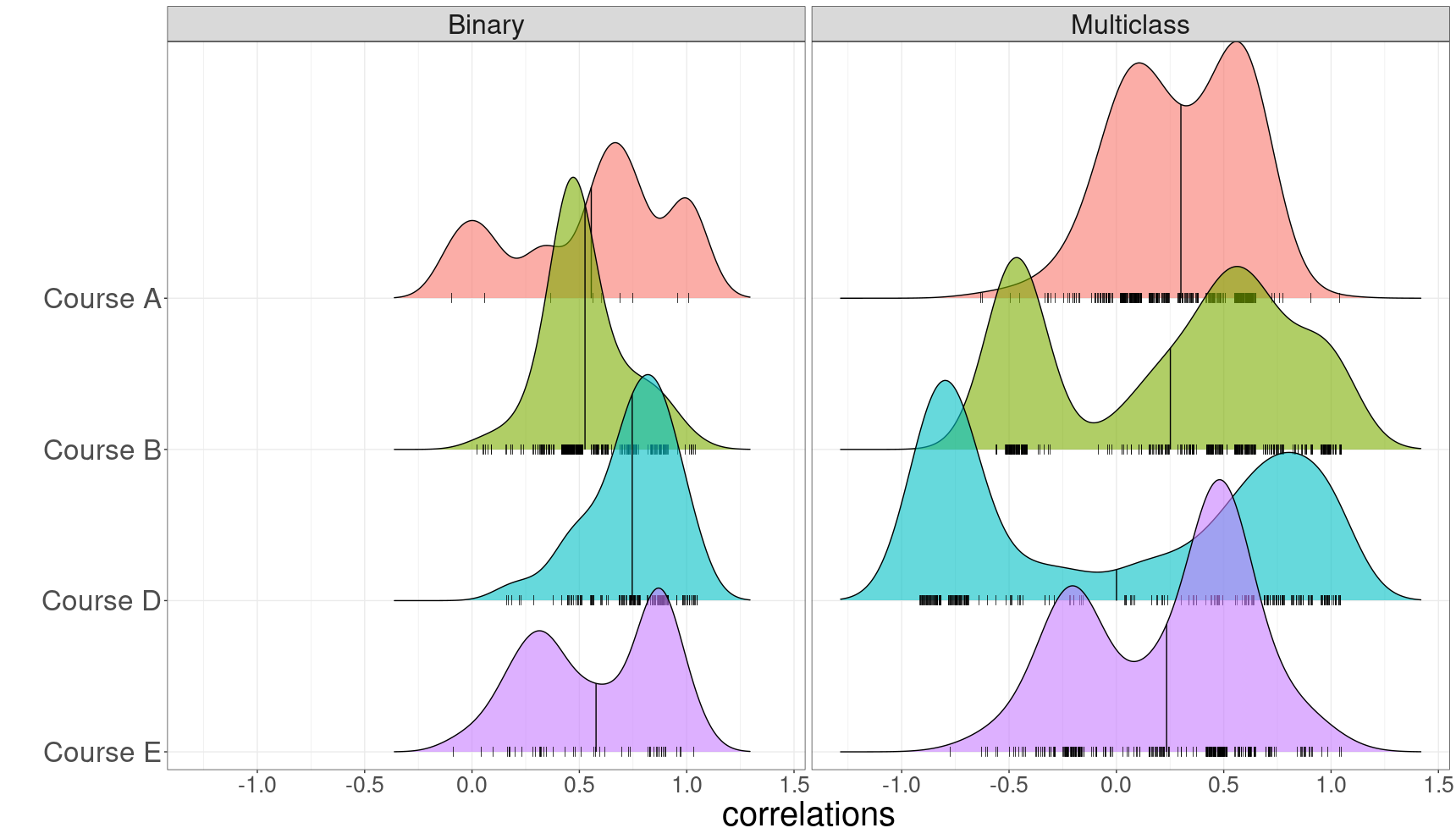}
    \caption{Variable Importance Order Similarity in the Rashomon set}
    \label{fig:kendall}
\end{figure}

\noindent \textbf{The distribution of the VIOD for the tasks and courses.} Figure~\ref{fig:kendall} presents the distribution of correlation values for binary and multiclass models across the courses. The x-axis represents Kendall's $\tau$ correlation coefficients, which measure the similarity between the variable importance order of the models in the Rashomon set and the reference model. The closer the correlation is to $1$ indicating the higher the similarity. Negative correlations, on the other hand, suggest that the variable importance order of the models in the Rashomon set is inversely related to the variable importance order of the reference model. The distributions reveal distinct patterns between binary and multiclass models. The binary models exhibit higher similarity for most courses, with distributions skewed toward positive values. In particular, the binary models for Courses B and D display unimodal distributions concentrated around high positive correlations, indicating that the models in the Rashomon set are similar to the reference model in terms of variable importance order. 

In contrast, the correlation distributions for multiclass models tend to be more spread out, suggesting greater variability in their variable importance order similarity. Notably, the multiclass models for Course A and Course E exhibit bimodal distributions, indicating that while some models achieve high positive correlations, others show weaker or near-zero correlations with the reference model. Binary models generally achieve stronger correlations than multiclass models, suggesting they may more effectively capture simpler patterns. Courses B and D stand out for their tightly concentrated binary model distributions, reflecting more stable and consistent similarities. On the other hand, Course A shows a wider spread of correlations in binary and multiclass settings, with the multiclass models demonstrating particularly diverse similarity levels, as evidenced by the bimodal shape. Course E also displays notable variability in terms of similarity, especially in the multiclass setting, where the distribution extends from low to high positive correlations, indicating that the performance of individual models varies substantially. 

%Figure~\ref{fig:kendall} shows the distributions of variable importance order similarity between the reference model and the other models in the Rashomon set for each course. The similarity between the variable importance order is measured using the Kendall correlation coefficient, which takes the values between $-1$ and $1$. The higher values represent the higher similarity, and vice versa. The vertical lines in the distribution representations indicate the mean similarity. For the binary response variable, the mean similarity value is around 0.5 for the courses, it is lower than $0.3$ for the multiclass response variable. It also takes the values $0$ or $1$ means some models have the same or quite different variable importance order similarity with the reference model in the Rashomon set.  

%%%%%%%%%%%%%%%%%%%%%%%%%%%%%%%%%%%%%%%%%%%%%%%%%%%
% NOTES
\section{Discussion}
In this study, we analysed the data from the OULAD dataset by applying the approach of \cite{rizvi_et_al_2019}. In contrast to the original research, we aim to highlight the importance of choosing a machine learning model for studied phenomena and promote using multiple algorithms to construct a set of models providing more insights from various "points of view". 

At first, we examined the predictive accuracies of the machine learning models set consisting of decision trees, random forests, lightGBM, and XGBoost tuned by Bayesian optimisation and Random Search. The formed set of models contained 424 models with overall accuracy ranging from approximately 0.62 to 0.80 for binary classification (\texttt{Pass} / \texttt{Fail} case) and from approximately 0.52 to 0.75 for the case of multiclass classification (\texttt{Distinct} / \texttt{Pass} / \texttt{Fail} case). When the model space is reduced to the Rashomon set of models with parameter $ \epsilon = 0.05 $, the overall classification accuracy is increased approximately by 2 to 6 \%. Further highlighting that model selection is a vital part of machine-learning modelling. The resulting accuracy of the models in the Rashomon set is slightly higher for the case of binary classification and approximately similar for the case of multiclass classification. 

The results also indicate that the "predictive value" of the demographic data in the OULAD dataset has been exceeded because even with the set of models, the overall accuracy is not increasing. Thus, the information contained within variables towards the target variable is probably exhausted. This concludes the answer to RQ1: How does the choice of training algorithm influence the predictive performance of the final model?

When further analysing the outcomes of the variable importance (RQ2: Are there any differences between the multiclass and binary classification tasks in terms of variable importance?), it can be observed that results highlight a distinction between binary and multiclass models. Binary models display more consistent variable importance, suggesting that the relationships between variables and outcomes are more straightforward when predicting binary outcomes. On the other hand, multiclass models introduce more variability, as different models may capture varying aspects of the data to optimise performance across multiple outcomes. This variability in multiclass setups suggests that it is more challenging to identify stable patterns across variables when the target variable has more classes. To conclude, the higher VIOD values in multiclass models emphasise the added complexity in predicting multiple outcomes, while the lower VIOD values in binary models reflect more reliable identification of important variables. The consistently high VIOD values observed for Course D across binary and multiclass setups indicate that student success in this course is hard to predict, potentially due to complex underlying relationships in the data. These findings offer valuable insights for educators and researchers, helping them understand the stability and reliability of variable importance rankings across different prediction frameworks. The results suggest that binary models yield more consistent higher correlation values, while multiclass models exhibit variability across courses. The bimodal distributions in the multiclass settings, particularly for Courses A and E, highlight the challenge of maintaining stable similarity when predicting multiple classes. These results underscore the importance of selecting appropriate models for each course and response type to achieve reliable information about the importance of the variables. 

Finally, we can observe that overall models in the Rashomon set suggest that the \texttt{imd\_band} and \texttt{highest\_education} are vital over the multiple courses. This importance is stronger for the binary models. The only exception is course D, where the number of important models and the accuracy impact are limited, which may explain the lower performance observed for this course. These results are also reflected by the highest VOID score of $-0.866$. The findings diverge from the original research, where the most important demographical characteristics were \texttt{region} and \texttt{imd\_band} followed by the \texttt{highest\_education} and \texttt{age\_band}. This shows that more model decisions are based on a slightly different set of demographical features than the one decision tree model. It is well-known that decision tree algorithms are sensitive to the input data \citep{zhou2012}, and resulting models can be significantly different.  

Our findings align with some of the previous research \citep{kizilcec2017self, richardson2009academic}, where the learner's previous education plays a significant role in academic outcomes prediction. However, the results contradict findings from the original study \citep{rizvi_et_al_2019} and others \citep{Wladis2015285,guo2014} further suggesting the influence of model choice on the variable importance (RQ3: How does the choice of models affect the variable importance?). 

Finally, it needs to be noted that even with the use of Rashomon sets for predictive modelling or estimation of the variable importance, the generalisation of the results needs to be done with caution \citep{Gašević201668}, and the models need to adapt to the individual context of the particular machine learning task. 

\section{Conclusions}

This paper explored the influence and importance of proper model selection for particular educational data. We experimented similarly to one of the well-established and highly reliable research studies from the past \citep{rizvi_et_al_2019}, and the results of our study partially matched with the previous research. To address the stable variable importance estimation, we employed the methods to develop the so-called Rashomon set of machine learning models[], which matches the reality within the accuracy range (set of equally well-performing models). The chosen approach addresses the old dilemma of all researchers working with machine learning models: "Which algorithm will likely work with my data?" \citep{Rudin2024} It has been shown that the choice of model is highly dependent on the level of noise in the dataset. In the case of educational data, noisy processes are often the case. Thus, the Rashomon effect is also strongly manifested within the training step of the machine learning modelling step. It is vital to address this matter when the interpretability of the model is in play. The Rashomon effect changes how researchers should approach the modelling of the phenomena and unlocks a set of tools for reliable, interpretable, and trustful machine learning models. Our study has several limitations: 1) the OULAD dataset, which contains demographic data with significantly reduced information due to the anonymisation process; 2) demographic variables have limited predictive value compared to the VLE-clickstream data in the context of OULAD; 3) the study used only a subset of the OULAD dataset (similar to \cite{rizvi_et_al_2019}), which might limit the generalisation of the results. 
We believe that the Rashomon effect will influence educational data mining research since it has a proven influence on the interpretability and stability of modelling results. Thus, it is critical to introduce the phenomena to a broader audience of scholars to address this issue, which changes the fundamental paradigm of how the machine learning experiment is conducted and algorithms are formulated.

%%%%%%%%%%%%%%%%%%%%%%%%%%%%%%%%%%%%%%%%%%%%%%%%%%%
\section*{Acknowledgments}
The work in this paper is supported by $<$anonymous grant agency$>$, grant no. $<$anonymous number$>$.

%Examples of a figure and a table are given in Figure~\ref{fig:example} and Table~\ref{tab:1}.

\section*{Declaration of Generative AI Software Tools in the Writing Process}

During the preparation of this work, the authors used ChatGPT 4.0 in Sections 3 and 4 to clarify the flow of the paper and Qwen2.5 to find synonyms for multiple words and phrases in the paper and to summarise the paper text to draft the first version of the abstract. After using this tool/service, the authors reviewed and edited the content as needed and take full responsibility for the content of the publication.

% REMOVE NOCITE OR IT WILL LIST EVERYTHING IN YOUR DATABASE AS A REFERENCE
%\nocite{*}

\bibliographystyle{acmtrans}
\bibliography{main}

\begin{thebibliography}{}

\bibitem[\protect\citeauthoryear{Abdelrahman, Wang, and Nunes}{Abdelrahman
  et~al\mbox{.}}{2023}]{abdelrahman}
{\sc Abdelrahman, G.}, {\sc Wang, Q.}, {\sc and} {\sc Nunes, B.} 2023.
\newblock Knowledge tracing: A survey.
\newblock {\em ACM Computing Surveys\/}.
\newblock Query date: 2024-12-01 10:01:18.

\bibitem[\protect\citeauthoryear{Abyaa, Idrissi, and Bennani}{Abyaa
  et~al\mbox{.}}{2019}]{Abyaa2019}
{\sc Abyaa, A.}, {\sc Idrissi, M.~K.}, {\sc and} {\sc Bennani, S.} 2019.
\newblock Learner modelling: systematic review of the literature from the last
  5 years.
\newblock {\em Educational Technology Research …\/}.
\newblock Query date: 2024-12-01 09:57:18.

\bibitem[\protect\citeauthoryear{Arnold and Pistilli}{Arnold and
  Pistilli}{2012a}]{arnold2012course}
{\sc Arnold, K.~E.} {\sc and} {\sc Pistilli, M.~D.} 2012a.
\newblock Course signals at purdue: Using learning analytics to increase
  student success.
\newblock In {\em Proceedings of the 2nd international conference on learning
  analytics and knowledge}. 267--270.

\bibitem[\protect\citeauthoryear{Arnold and Pistilli}{Arnold and
  Pistilli}{2012b}]{Arnold2012}
{\sc Arnold, K.~E.} {\sc and} {\sc Pistilli, M.~D.} 2012b.
\newblock {Course Signals at Purdue: Using Learning Analytics to Increase
  Student Success}.
\newblock {\em Proceedings of the 2nd International Conference on Learning
  Analytics and Knowledge\/}~May, 267--270.

\bibitem[\protect\citeauthoryear{Balaji, Alelyani, Qahmash, and Mohana}{Balaji
  et~al\mbox{.}}{2021}]{balaji2021contributions}
{\sc Balaji, P.}, {\sc Alelyani, S.}, {\sc Qahmash, A.}, {\sc and} {\sc Mohana,
  M.} 2021.
\newblock Contributions of machine learning models towards student academic
  performance prediction: a systematic review.
\newblock {\em Applied Sciences\/}~{\em 11,\/}~21, 10007.

\bibitem[\protect\citeauthoryear{Bayeck and Choi}{Bayeck and
  Choi}{2018}]{bayeck2018influence}
{\sc Bayeck, R.~Y.} {\sc and} {\sc Choi, J.} 2018.
\newblock The influence of national culture on educational videos: The case of
  moocs.
\newblock {\em International Review of Research in Open and Distributed
  Learning\/}~{\em 19,\/}~1.

\bibitem[\protect\citeauthoryear{Bergstra and Bengio}{Bergstra and
  Bengio}{2012}]{bergstra_et_al_2012}
{\sc Bergstra, J.} {\sc and} {\sc Bengio, Y.} 2012.
\newblock Random search for hyper-parameter optimization.
\newblock {\em Journal of Machine Learning Research\/}~{\em 13,\/}~2, 281--305.

\bibitem[\protect\citeauthoryear{Biecek, Baniecki, Krzyzi{\'n}ski, and
  Cook}{Biecek et~al\mbox{.}}{2024}]{biecek_et_al_2024}
{\sc Biecek, P.}, {\sc Baniecki, H.}, {\sc Krzyzi{\'n}ski, M.}, {\sc and} {\sc
  Cook, D.} 2024.
\newblock Performance is not enough: the story told by a rashomon quartet.
\newblock {\em Journal of Computational and Graphical
  Statistics\/}~just-accepted, 1--6.

\bibitem[\protect\citeauthoryear{Boyte-Eckis, Minadeo, Bailey, and
  Bailey}{Boyte-Eckis et~al\mbox{.}}{2018}]{boyte2018age}
{\sc Boyte-Eckis, L.}, {\sc Minadeo, D.~F.}, {\sc Bailey, S.~S.}, {\sc and}
  {\sc Bailey, W.~C.} 2018.
\newblock Age, gender, and race as predictors of opting for a midterm retest: A
  statistical analysis of online economics students.
\newblock {\em Journal of Business Diversity\/}~{\em 18,\/}~1.

\bibitem[\protect\citeauthoryear{Breiman}{Breiman}{2001a}]{breiman_2001}
{\sc Breiman, L.} 2001a.
\newblock Random forests.
\newblock {\em Machine learning\/}~{\em 45}, 5--32.

\bibitem[\protect\citeauthoryear{Breiman}{Breiman}{2001b}]{breiman2001statistical}
{\sc Breiman, L.} 2001b.
\newblock Statistical modeling: The two cultures (with comments and a rejoinder
  by the author).
\newblock {\em Statistical science\/}~{\em 16,\/}~3, 199--231.

\bibitem[\protect\citeauthoryear{Breiman, Friedman, Olshen, and Stone}{Breiman
  et~al\mbox{.}}{1984}]{breiman_et_al_1984}
{\sc Breiman, L.}, {\sc Friedman, J.}, {\sc Olshen, R.}, {\sc and} {\sc Stone,
  C.} 1984.
\newblock {\em Classification and Regression Trees}.
\newblock Wadsworth International Group.

\bibitem[\protect\citeauthoryear{Cavus and Biecek}{Cavus and
  Biecek}{2024}]{cavus_and_biecek_2024}
{\sc Cavus, M.} {\sc and} {\sc Biecek, P.} 2024.
\newblock An experimental study on the rashomon effect of balancing methods in
  imbalanced classification.
\newblock {\em arXiv preprint arXiv:2405.01557\/}.

\bibitem[\protect\citeauthoryear{Chen and Guestrin}{Chen and
  Guestrin}{2016}]{chen_and_carlos_2016}
{\sc Chen, T.} {\sc and} {\sc Guestrin, C.} 2016.
\newblock Xgboost: A scalable tree boosting system.
\newblock In {\em Proceedings of the 22nd acm sigkdd international conference
  on knowledge discovery and data mining}. 785--794.

\bibitem[\protect\citeauthoryear{Davis, Chen, Hauff, and Houben}{Davis
  et~al\mbox{.}}{2016}]{davis2016gauging}
{\sc Davis, D.}, {\sc Chen, G.}, {\sc Hauff, C.}, {\sc and} {\sc Houben, G.-J.}
  2016.
\newblock Gauging mooc learners' adherence to the designed learning path.
\newblock {\em International Educational Data Mining Society\/}.

\bibitem[\protect\citeauthoryear{Dong and Rudin}{Dong and
  Rudin}{2020}]{dong_and_rudin_2020}
{\sc Dong, J.} {\sc and} {\sc Rudin, C.} 2020.
\newblock Exploring the cloud of variable importance for the set of all good
  models.
\newblock {\em Nature Machine Intelligence\/}~{\em 2,\/}~12, 810--824.

\bibitem[\protect\citeauthoryear{Donnelly, Katta, Rudin, and Browne}{Donnelly
  et~al\mbox{.}}{2023}]{donnelly_et_al_2023}
{\sc Donnelly, J.}, {\sc Katta, S.}, {\sc Rudin, C.}, {\sc and} {\sc Browne,
  E.} 2023.
\newblock The rashomon importance distribution: Getting rid of unstable, single
  model-based variable importance.
\newblock {\em Advances in Neural Information Processing Systems\/}~{\em 36},
  6267--6279.

\bibitem[\protect\citeauthoryear{Fisher, Rudin, and Dominici}{Fisher
  et~al\mbox{.}}{2019}]{fisher_et_al_2019}
{\sc Fisher, A.}, {\sc Rudin, C.}, {\sc and} {\sc Dominici, F.} 2019.
\newblock All models are wrong, but many are useful: Learning a variable's
  importance by studying an entire class of prediction models simultaneously.
\newblock {\em Journal of Machine Learning Research\/}~{\em 20,\/}~177, 1--81.

\bibitem[\protect\citeauthoryear{Guo and Reinecke}{Guo and
  Reinecke}{2014}]{guo2014}
{\sc Guo, P.~J.} {\sc and} {\sc Reinecke, K.} 2014.
\newblock Demographic differences in how students navigate through moocs.
\newblock In {\em Proceedings of the First ACM Conference on Learning @ Scale
  Conference}. L@S '14. Association for Computing Machinery, New York, NY, USA,
  21–30.

\bibitem[\protect\citeauthoryear{Harif and Kassimi}{Harif and
  Kassimi}{2024}]{harif2024predictive}
{\sc Harif, A.} {\sc and} {\sc Kassimi, M.~A.} 2024.
\newblock Predictive modeling of student performance using rfecv-rf for feature
  selection and machine learning techniques.
\newblock {\em International Journal of Advanced Computer Science and
  Applications (IJACSA)\/}~{\em 15,\/}~7.

\bibitem[\protect\citeauthoryear{Ke, Meng, Finley, Wang, Chen, Ma, Ye, and
  Liu}{Ke et~al\mbox{.}}{2017}]{ke_et_al_2017}
{\sc Ke, G.}, {\sc Meng, Q.}, {\sc Finley, T.}, {\sc Wang, T.}, {\sc Chen, W.},
  {\sc Ma, W.}, {\sc Ye, Q.}, {\sc and} {\sc Liu, T.-Y.} 2017.
\newblock Lightgbm: A highly efficient gradient boosting decision tree.
\newblock {\em Advances in neural information processing systems\/}~{\em 30}.

\bibitem[\protect\citeauthoryear{Kendall}{Kendall}{1938}]{kendall_1938}
{\sc Kendall, M.~G.} 1938.
\newblock A new measure of rank correlation.
\newblock {\em Biometrika\/}~{\em 30,\/}~1-2, 81--93.

\bibitem[\protect\citeauthoryear{Kennedy, Coffrin, de~Barba, and
  Corrin}{Kennedy et~al\mbox{.}}{2015}]{Kennedy2015}
{\sc Kennedy, G.}, {\sc Coffrin, C.}, {\sc de~Barba, P.}, {\sc and} {\sc
  Corrin, L.} 2015.
\newblock {Predicting success: how learners' prior knowledge, skills and
  activities predict MOOC performance}.
\newblock {\em Proceedings of the Fifth International Conference on Learning
  Analytics And Knowledge - LAK '15\/}, 136--140.

\bibitem[\protect\citeauthoryear{Kent, Boulton, and Williams}{Kent
  et~al\mbox{.}}{2017}]{Kent2017}
{\sc Kent, C.}, {\sc Boulton, C.~A.}, {\sc and} {\sc Williams, H.} 2017.
\newblock {Towards Measurement of the Relationship between Student Engagement
  and Learning Outcomes at a Bricks-and-Mortar University}.
\newblock {\em Joint Proceedings of the Sixth Multimodal Learning Analytics
  (MMLA) Workshop and the Second Cross-LAK Workshop (MMLA-CrossLAK)\/}~1828,
  4--14.

\bibitem[\protect\citeauthoryear{Kizilcec, P{\'e}rez-Sanagust{\'\i}n, and
  Maldonado}{Kizilcec et~al\mbox{.}}{2017}]{kizilcec2017self}
{\sc Kizilcec, R.~F.}, {\sc P{\'e}rez-Sanagust{\'\i}n, M.}, {\sc and} {\sc
  Maldonado, J.~J.} 2017.
\newblock Self-regulated learning strategies predict learner behavior and goal
  attainment in massive open online courses.
\newblock {\em Computers \& education\/}~{\em 104}, 18--33.

\bibitem[\protect\citeauthoryear{Klerkx, Verbert, and Duval}{Klerkx
  et~al\mbox{.}}{2017}]{klerkx2017learning}
{\sc Klerkx, J.}, {\sc Verbert, K.}, {\sc and} {\sc Duval, E.} 2017.
\newblock Learning analytics dashboards.
\newblock {\em Handbook of learning analytics\/}~{\em 1}, 143--150.

\bibitem[\protect\citeauthoryear{Kurosawa}{Kurosawa}{1950}]{kurosawa1950}
{\sc Kurosawa, A.} 1950.
\newblock {\em Rashomon}.
\newblock RKO Radio Pictures.

\bibitem[\protect\citeauthoryear{Kuzilek, Hlosta, Herrmannova, Zdrahal,
  Vaclavek, and Wolff}{Kuzilek et~al\mbox{.}}{2015}]{Kuzilek2015}
{\sc Kuzilek, J.}, {\sc Hlosta, M.}, {\sc Herrmannova, D.}, {\sc Zdrahal, Z.},
  {\sc Vaclavek, J.}, {\sc and} {\sc Wolff, A.} 2015.
\newblock {LAK15 Case Study 1: OU Analyse: Analysing At-Risk Students at The
  Open University - Learning Analytics Review}.
\newblock {\em Learning Analytics Review\/}.

\bibitem[\protect\citeauthoryear{Kuzilek, Hlosta, and Zdrahal}{Kuzilek
  et~al\mbox{.}}{2017a}]{Kuzilek2017}
{\sc Kuzilek, J.}, {\sc Hlosta, M.}, {\sc and} {\sc Zdrahal, Z.} 2017a.
\newblock {Open University Learning Analytics dataset}.
\newblock {\em Scientific Data\/}~{\em 4}, 170171.

\bibitem[\protect\citeauthoryear{Kuzilek, Hlosta, and Zdrahal}{Kuzilek
  et~al\mbox{.}}{2017b}]{kuzilek_et_al_2017}
{\sc Kuzilek, J.}, {\sc Hlosta, M.}, {\sc and} {\sc Zdrahal, Z.} 2017b.
\newblock Open university learning analytics dataset.
\newblock {\em Scientific data\/}~{\em 4,\/}~1, 1--8.

\bibitem[\protect\citeauthoryear{Molnar}{Molnar}{2020}]{molnar_et_al_2020}
{\sc Molnar, C.} 2020.
\newblock {\em Interpretable machine learning}.
\newblock Lulu. com.

\bibitem[\protect\citeauthoryear{Murphy}{Murphy}{2022}]{murphy_2022}
{\sc Murphy, K.~P.} 2022.
\newblock {\em Probabilistic Machine Learning: An introduction}.
\newblock MIT Press.

\bibitem[\protect\citeauthoryear{Papamitsiou and Economides}{Papamitsiou and
  Economides}{2014a}]{papamitsiou2014learning}
{\sc Papamitsiou, Z.} {\sc and} {\sc Economides, A.~A.} 2014a.
\newblock Learning analytics and educational data mining in practice: A
  systematic literature review of empirical evidence.
\newblock {\em Journal of Educational Technology \& Society\/}~{\em 17,\/}~4,
  49--64.

\bibitem[\protect\citeauthoryear{Papamitsiou and Economides}{Papamitsiou and
  Economides}{2014b}]{Papamitsiou2014}
{\sc Papamitsiou, Z.} {\sc and} {\sc Economides, A.~A.} 2014b.
\newblock {Learning Analytics and Educational Data Mining in Practice: A
  Systematic Literature Review of Empirical Evidence}.
\newblock {\em Educational Technology {\&} Society\/}~{\em 17,\/}~4, 49--64.

\bibitem[\protect\citeauthoryear{Raschka}{Raschka}{2018}]{Raschka2018ModelEM}
{\sc Raschka, S.} 2018.
\newblock Model evaluation, model selection, and algorithm selection in machine
  learning.
\newblock {\em ArXiv\/}~{\em abs/1811.12808}.

\bibitem[\protect\citeauthoryear{Richardson}{Richardson}{2009}]{richardson2009academic}
{\sc Richardson, J.~T.} 2009.
\newblock The academic attainment of students with disabilities in uk higher
  education.
\newblock {\em Studies in Higher Education\/}~{\em 34,\/}~2, 123--137.

\bibitem[\protect\citeauthoryear{Rizvi, Rienties, and Khoja}{Rizvi
  et~al\mbox{.}}{2019}]{rizvi_et_al_2019}
{\sc Rizvi, S.}, {\sc Rienties, B.}, {\sc and} {\sc Khoja, S.~A.} 2019.
\newblock The role of demographics in online learning; a decision tree based
  approach.
\newblock {\em Computers \& Education\/}~{\em 137}, 32--47.

\bibitem[\protect\citeauthoryear{Rodolfa, Salomon, Haynes, Mendieta, Larson,
  and Ghani}{Rodolfa et~al\mbox{.}}{2020}]{rodolfa_et_al_2020}
{\sc Rodolfa, K.~T.}, {\sc Salomon, E.}, {\sc Haynes, L.}, {\sc Mendieta,
  I.~H.}, {\sc Larson, J.}, {\sc and} {\sc Ghani, R.} 2020.
\newblock Case study: predictive fairness to reduce misdemeanor recidivism
  through social service interventions.
\newblock In {\em Proceedings of the 2020 Conference on Fairness,
  Accountability, and Transparency}. 142--153.

\bibitem[\protect\citeauthoryear{Romero, L{\'{o}}pez, Luna, and Ventura}{Romero
  et~al\mbox{.}}{2013}]{Romero2013}
{\sc Romero, C.}, {\sc L{\'{o}}pez, M.~I.}, {\sc Luna, J.~M.}, {\sc and} {\sc
  Ventura, S.} 2013.
\newblock {Predicting students' final performance from participation in on-line
  discussion forums}.
\newblock {\em Computers and Education\/}~{\em 68}, 458--472.

\bibitem[\protect\citeauthoryear{Ruczy{\'n}ski and Kozak}{Ruczy{\'n}ski and
  Kozak}{2024}]{ruczynski_and_kozak_2024}
{\sc Ruczy{\'n}ski, H.} {\sc and} {\sc Kozak, A.} 2024.
\newblock forester: A tree-based automl tool in r.
\newblock {\em arXiv preprint arXiv:2409.04789\/}.

\bibitem[\protect\citeauthoryear{Rudin, Chen, Chen, Huang, Semenova, and
  Zhong}{Rudin et~al\mbox{.}}{2022}]{rudin_et_al_2022}
{\sc Rudin, C.}, {\sc Chen, C.}, {\sc Chen, Z.}, {\sc Huang, H.}, {\sc
  Semenova, L.}, {\sc and} {\sc Zhong, C.} 2022.
\newblock Interpretable machine learning: Fundamental principles and 10 grand
  challenges.
\newblock {\em Statistic Surveys\/}~{\em 16}, 1--85.

\bibitem[\protect\citeauthoryear{Rudin, Zhong, Semenova, Seltzer, Parr, Liu,
  Katta, Donnelly, Chen, and Boner}{Rudin et~al\mbox{.}}{2024}]{Rudin2024}
{\sc Rudin, C.}, {\sc Zhong, C.}, {\sc Semenova, L.}, {\sc Seltzer, M.}, {\sc
  Parr, R.}, {\sc Liu, J.}, {\sc Katta, S.}, {\sc Donnelly, J.}, {\sc Chen,
  H.}, {\sc and} {\sc Boner, Z.} 2024.
\newblock Amazing things come from having many good models.
\newblock In {\em Proceedings of the International Conference on Machine
  Learning (ICML)}.

\bibitem[\protect\citeauthoryear{Semenova, Chen, Parr, and Rudin}{Semenova
  et~al\mbox{.}}{2023}]{semenova2023}
{\sc Semenova, L.}, {\sc Chen, H.}, {\sc Parr, R.}, {\sc and} {\sc Rudin, C.}
  2023.
\newblock A path to simpler models starts with noise.
\newblock In {\em Proceedings of the 37th International Conference on Neural
  Information Processing Systems}. NIPS '23. Curran Associates Inc., Red Hook,
  NY, USA.

\bibitem[\protect\citeauthoryear{Sharabiani, Karim, Sharabiani, Atanasov, and
  Darabi}{Sharabiani et~al\mbox{.}}{2014}]{Sharabiani2014}
{\sc Sharabiani, A.}, {\sc Karim, F.}, {\sc Sharabiani, A.}, {\sc Atanasov,
  M.}, {\sc and} {\sc Darabi, H.} 2014.
\newblock {An enhanced bayesian network model for prediction of students'
  academic performance in engineering programs}.
\newblock In {\em 2014 IEEE Global Engineering Education Conference (EDUCON)}.
  832--837.

\bibitem[\protect\citeauthoryear{Shehata and Arnold}{Shehata and
  Arnold}{2015}]{Shehata2015}
{\sc Shehata, S.} {\sc and} {\sc Arnold, K.~E.} 2015.
\newblock {Measuring student success using predictive engine}.
\newblock {\em Proceedings of the Fifth International Conference on Learning
  Analytics And Knowledge - LAK '15\/}, 416--417.

\bibitem[\protect\citeauthoryear{Siemens and Baker}{Siemens and
  Baker}{2012}]{siemens2012learning}
{\sc Siemens, G.} {\sc and} {\sc Baker, R. S.~d.} 2012.
\newblock Learning analytics and educational data mining: towards communication
  and collaboration.
\newblock In {\em Proceedings of the 2nd international conference on learning
  analytics and knowledge}. 252--254.

\bibitem[\protect\citeauthoryear{Snoek, Larochelle, and Adams}{Snoek
  et~al\mbox{.}}{2012}]{snoek_et_al_2012}
{\sc Snoek, J.}, {\sc Larochelle, H.}, {\sc and} {\sc Adams, R.~P.} 2012.
\newblock Practical bayesian optimization of machine learning algorithms.
\newblock In {\em Advances in Neural Information Processing Systems}. Vol.~25.
  2951--2959.

\bibitem[\protect\citeauthoryear{Wickham, {\c{C}}etinkaya-Rundel, and
  Grolemund}{Wickham et~al\mbox{.}}{2023}]{wickham2023}
{\sc Wickham, H.}, {\sc {\c{C}}etinkaya-Rundel, M.}, {\sc and} {\sc Grolemund,
  G.} 2023.
\newblock {\em R for data science}.
\newblock " O'Reilly Media, Inc.".

\bibitem[\protect\citeauthoryear{Wladis, Hachey, and Conway}{Wladis
  et~al\mbox{.}}{2014}]{Wladis2014}
{\sc Wladis, C.}, {\sc Hachey, A.~C.}, {\sc and} {\sc Conway, K.} 2014.
\newblock {An investigation of course-level factors as predictors of online
  STEM course outcomes}.
\newblock {\em Computers and Education\/}~{\em 77}, 145--150.

\bibitem[\protect\citeauthoryear{Wladis, Hachey, and Conway}{Wladis
  et~al\mbox{.}}{2015}]{Wladis2015285}
{\sc Wladis, C.}, {\sc Hachey, A.~C.}, {\sc and} {\sc Conway, K.} 2015.
\newblock Which stem majors enroll in online courses, and why should we care?
  the impact of ethnicity, gender, and non-traditional student characteristics.
\newblock {\em Computers and Education\/}~{\em 87}, 285 – 308.
\newblock Cited by: 31; All Open Access, Bronze Open Access.

\bibitem[\protect\citeauthoryear{Zhou}{Zhou}{2012}]{zhou2012}
{\sc Zhou, Z.-H.} 2012.
\newblock {\em Ensemble Methods: Foundations and Algorithms\/}, 1st ed.
\newblock Chapman \& Hall/CRC.

\end{thebibliography}

\end{document}